\documentclass[aip,amsmath,amssymb,reprint,]{revtex4-1}

\usepackage{graphicx}% Include figure files
\usepackage{dcolumn}% Align table columns on decimal point
\usepackage{bm}% bold math
\usepackage[utf8]{inputenc}
\usepackage[T1]{fontenc}
\usepackage{mathptmx}
\usepackage{etoolbox}

\makeatletter
\def\@email#1#2{%
 \endgroup
 \patchcmd{\titleblock@produce}
  {\frontmatter@RRAPformat}
  {\frontmatter@RRAPformat{\produce@RRAP{*#1\href{mailto:#2}{#2}}}\frontmatter@RRAPformat}
  {}{}
}%
\makeatother
\begin{document}

\preprint{AIP/123-QED}

\title[Electronic and Optical Properties of Ta$_2$NiSe$_5$ Monolayer: A First-principles Study]{Electronic and Optical Properties of Ta$_2$NiSe$_5$ Monolayer: A First-principles Study}

\author{Miaomiao Guo}
\author{Yuanchang Li}
 \email{yuancli@bit.edu.cn.}
\affiliation{Key Lab of advanced optoelectronic quantum architecture and measurement (MOE), and Advanced Research Institute of Multidisciplinary Science, Beijing Institute of Technology, Beijing 100081, China.}

\date{\today}% It is always \today, today,
             %  but any date may be explicitly specified

\begin{abstract}
%\section{ABSTRACT}
The crystal structure, stability, electronic and optical properties of the Ta$_2$NiSe$_5$ monolayer have been investigated using first-principles calculations in combination with the Bethe-Salpeter equation. The results show that it is feasible to directly exfoliate a Ta$_2$NiSe$_5$ monolayer from the low-temperature monoclinic phase. The monolayer is stable and behaves as a normal narrow-gap semiconductor with neither spontaneous excitons nor non-trivial topology. Despite the quasi-particle and optical gaps of only 266 and 200 meV, respectively, its optically-active exciton has a binding energy up to 66 meV and can exist at room temperature. This makes it valuable for applications in infrared photodetection, especially its inherent in-plane anisotropy adds to its value in polarization sensing. It is also found that the inclusion of spin-orbit coupling is theoretically necessary to properly elucidate the optical and excitonic properties of monolayer.
\end{abstract}

\maketitle

%\section{\label{sec:level1}Introduction}
Two-dimensional (2D) materials have become one of the most active research topics in condensed matter physics, nanotechnology and materials science due to their unique structures and physical properties. Since the exfoliation of graphene in 2004\cite{Novoselov 2024}, hundreds of 2D materials, including black phosphorus, Bi$_2$Se$_3$ family, transition-metal dichalcogenides, and MXenes, have been obtained so far, which show great potentials for applications\cite{Wu 2014, Zhang 2015, Xue 2017, Acerce 2015} in electronic/optoelectronic devices, ultrafast laser generation, sensors, catalysis, and energy storage. Among 2D materials, monolayers that can be exfoliated from layered bulk are particularly interesting, not only because they are relatively easy to prepare, but also because they can be reassembled into a variety of van der Waals heterostructures thereby realizing complex and extraordinary physical phenomena\cite{Zhang 2022,Wang 2023}. In recent years, 2D ternary compounds have attracted more and more attention. These materials add a degree of freedom compared to binary compounds, thus allowing for emerging physics through synergistic effects, e.g., the quantum anomalous Hall effect and axion electrodynamics in MnBi$_2$Te$_4$\cite{Li 2019}.

Ternary layered Ta$_2$NiSe$_5$ is first prepared and characterized as a semiconductor with a gap of 0.36 eV by Sunshine and Ibers in 1985\cite{Sunshine 1985}. Over the next two decades, Ta$_2$NiSe$_5$ has not attracted much attention. This continues until 2009, when Wakisaka et al. invoke the excitonic-insulator scenario to interpret the gap opening during an orthorhombic-monoclinic phase transition at 328 K\cite{Wakisaka 2009}. Their angle-resolved photoemission spectrum studies reveal an extremely flat dispersion at the valence-band top, which is a distinctive feature of the excitonic-insulator ground state. This finding has sparked a large number of subsequent studies into its possible excitonic-insulator phase\cite{Kaneko 2013,Chatterjee,Mazza 2020,Watson 2020,Subedi 2020,Kim 2020,Windgatter 2021,Baldini 2023,Baldini 2023,Kim 2016,Liu 2021}. Since no charge-density-wave is formed during the phase transition, it was initially thought that Ta$_2$NiSe$_5$ could avoid the difficulty of distinguishing from the lattice instability that occurs in TiSe$_2$\cite{Jiang 2018}. However, recent theoretical and experimental studies have pointed out that the phase transition may be alternatively driven by a softening of a $B_{2g}$ zone-center phonon mode\cite{Subedi 2020,Kim 2020}. This has prevented Ta$_2$NiSe$_5$ from escaping the debate of whether the transition is originated from an excitonic-insulator or band-type Jahn-Teller mechanism\cite{Windgatter 2021, Baldini 2023}. By definition, excitonic insulators are crystals that have spontaneous excitons at 0 K\cite{Halperin 1968}. As such, in addition to whether or not the excitons drive the phase transition, whether or not there are spontaneous excitons in the low-temperature phase is another issue of interest\cite{Windgatter 2021}. It is worth noting that the spontaneous excitons depend only on the relative magnitudes of the single-particle energy gap and exciton binding energy of the crystal itself\cite{Halperin 1968}, which is essentially a different issue from the exciton-driven phase transition. Even if the orthorhombic-monoclinic phase transition of Ta$_2$NiSe$_5$ is exclusively exciton-driven, the presence of spontaneous excitons in the low-temperature phase cannot be guaranteed. The reason for this is that the orthorhombic-monoclinic structural distortions require electron/exciton-phonon coupling to produce phonons. This process transfers energy from excitons to phonons, thus reducing the number of excitons. If all excitons transfer energy to phonons, no spontaneous excitons will survive in the final monoclinic structure (excitonic order disappears and only structural order remains). Otherwise, if only some of the excitons transfer energy to phonons, there will be spontaneous excitons surviving in the final monoclinic structure, resulting in both excitonic and structural orders\cite{Chatterjee,Kaneko 2013}.

In 2015, Tan et al. obtain ultrathin Ta$_2$NiSe$_5$ nanosheets with 2$\sim$5 layers using electrochemical lithium intercalation-assisted exfoliation\cite{Tan 2015}, which naturally opens up some opportunities. As the thickness decreases, the electron-hole screening interaction is significantly weakened, leading to an increase in the exciton binding energy, which undoubtedly favors excitonic instability\cite{Jiang 2018, Jiang 2017}. For example, bulk ZrTe$_2$ is metallic with a negative gap of 0.5 eV, whereas thinning to the monolayer results in a charge-density-wave state associated with the excitonic insulator\cite{Gao 2023, Song 2023}. In addition, multilayer 2D Ta$_2$NiSe$_5$ are fabricated as photodetectors, showing excellent flexibility and photodetection performance\cite{LiAFM 2016, Zhang 2021, GuoT 2023}. Nevertheless, the electronic structure of 2D Ta$_2$NiSe$_5$ has not yet been studied in theory.

In this work, we investigate the structural, electronic and excitonic properties of monolayer Ta$_2$NiSe$_5$ in the monoclinic structure using first-principles calculations in combination with the Bethe-Salpeter equation (BSE), focusing on the presence or absence of spontaneous excitons and leaving the phase transition issue for the future. Energetic calculations show that it is highly possible to exfoliate a Ta$_2$NiSe$_5$ monolayer directly from its layered low-temperature monoclinic phase. Phonon spectrum calculations show that the monolayer is dynamically stable. Electronic structure calculations show that the monolayer is a direct-gap semiconductor with a quasi-particle gap of 266 meV. Solving the BSE yields an optical gap of 200 meV and an exciton binding energy of 66 meV. We also examine the effect of spin-orbit coupling and find that it only increases the gap by about 30\% without causing gap closure and reopening. These results point out that (1) the inter-layer interactions in Ta$_2$NiSe$_5$ are weak, and (2) the monolayer Ta$_2$NiSe$_5$ is a narrow-gap semiconductor with neither spontaneous excitons nor non-trivial topology.

%\section{\label{sec:level1}Methodology and models}
Our geometric and electronic calculations were performed using the Vienna \emph{ab initio} simulation package (VASP)\cite{vasp 1996} within the Perdew-Burke-Ernzerhof (PBE) exchange-correlation functional\cite{PBE 1996}. Spin-orbit coupling (SOC) is included, taking into account the presence of the heavy element Ta. Projector augmented wave (PAW)\cite{PAW 1994} method was used with an energy cutoff of 300 eV. A vacuum layer of 17 \AA\ was used to avoid spurious interactions between two neighboring images. A $15 \times 1 \times 3$ $k$-grid was used for sampling the Brillouin zone. Fully structural relaxation including atomic positions, cell shape, and cell volume was carried out until the residual force on each atom is less than 0.01 eV/\AA. To cure the gap underestimation of PBE, single-shot $G_0W_0$ calculations were performed for quasi-particle band structure\cite{G0W0 1986}. In a balance between our currently affordable computational cost and accuracy, the $G_0W_0$ calculations used an energy cutoff of 150 eV and a total of 576 and 384 bands with and without SOC, respectively. Excitonic properties were obtained by solving the BSE\cite{BSE 2000} on top of the $G_0W_0$, with 12 valence and 12 conduction bands for building the Hamiltonian. Phonon spectrum was calculated within density functional perturbation theory with the cutoff energy of 300 eV, the $k$-grid of $9 \times 1 \times 2$ and the $q$-grid of $9 \times 1 \times 9$.

%\section{\label{sec:level1}Results and discussion}
%\subsection{\label{sec:level2}Structure and energetics}

%fig01 Structure
\begin{figure}[htbp]
\includegraphics[width=0.95\columnwidth]{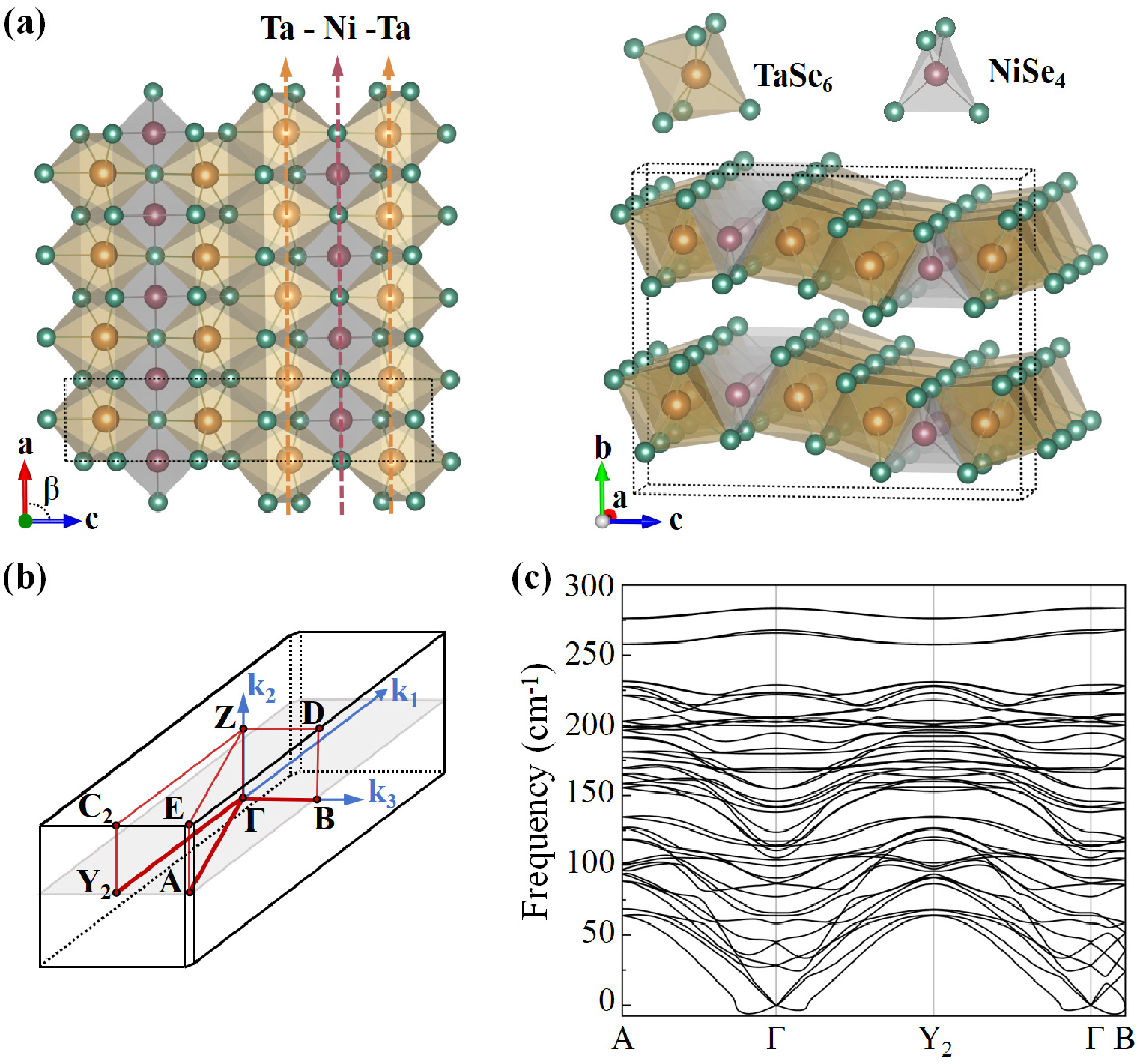}
\caption{\label{fig:fig1} (a) Top and side views of the low-temperature monoclinic Ta$_2$NiSe$_5$, as well as the Ta/Ni-centered local octahedron/tetrahedron. The arrows guide the direction of the Ta-Ni-Ta trimetal chain to highlight the quasi-one-dimensional structure. Black dashed boxes indicate the unit cells used for calculations. (b) The corresponding Brillouin zone, with $\Gamma$$Y_2$ denoting the direction along the Ta-Ni-Ta trimetal chain and $\Gamma$$B$ denoting the direction perpendicular to the Ta-Ni-Ta trimetal chain. (c) The calculated phonon spectrum of monolayer Ta$_2$NiSe$_5$.}
\end{figure}

Figures 1(a) and 1(b) shows the geometric configuration and corresponding Brillouin zone of bulk Ta$_2$NiSe$_5$. It is a layered crystal stacked by van der Waals interactions along the $b$-axis. Each layer is 3-atom thick and the Ta/Ni is located at the octahedral/tetrahedral centers formed by the surrounding Se, constituting the TaSe$_6$/NiSe$_4$ units. TaSe$_6$ and NiSe$_4$ extend along the $a$-axis to form chains, and the Ta-Ni-Ta triple chain repeats along the $c$-axis to form a quasi-one-dimensional structure. It belongs to the monoclinic system with space group C$_2$/c and experimental crystal parameters\cite{Nakano 2018} of $a$ = 3.49 \AA, $b$ = 12.81 \AA, $c$ = 15.65 \AA, and $\alpha$ = $\gamma$ = 90$^\circ$, $\beta$ = 90.53$^\circ$.

After a full relaxation of the monolayer directly peeled from bulk Ta$_2$NiSe$_5$, we obtain its optimized crystal parameters of $a$ = 3.51 \AA, $c$ = 15.77 \AA, and $\beta$ = 89.45$^\circ$. Compared with those of the bulk, there is a slight increase in $a$ and $c$, and a slight decrease in $\beta$. Table I gives the optimized atomic positions in the monolayer. They change only slightly compared to those in the bulk, with the largest shift of 0.15 \AA\ occurring in Se(1) along the $c$-axis. This implies very weak inter-layer interactions in the bulk Ta$_2$NiSe$_5$. To assess the structural stability of the monolayer, we have calculated its phonon spectrum, as shown in Fig. 1(c). The absence of imaginary frequencies indicates that the monolayer is dynamically stable. Note that tiny imaginary frequencies near the $\Gamma$ point also appear in the phonon spectra of other 2D systems\cite{Gao 2017, Sahin 2009}, which are not a true sign of structural instability. Its occurrence is usually due to the sensitivity to numerical accuracy (\emph{e.g.}, mesh size) caused by the rapid decay of the interatomic
forces\cite{Gao 2017, Sahin 2009}.

%Table 1
\begin{table}[htbp]
	\caption{Optimized atomic positions in the Ta$_2$NiSe$_5$ monolayer.}
	\centering
		\begin{tabular}{c c c c c }
			\hline
			\hline
      \multicolumn{1}{c}{ }&\multicolumn{4}{c}{Wyckoff Positions}\\
      \hline
      \makebox[0.085\textwidth][c]{Atom}&\makebox[0.085\textwidth][c]{Site}&\makebox[0.085\textwidth][c]{x}&\makebox[0.085\textwidth][c]{y}&\makebox[0.085\textwidth][c]{z}\\
     \hline
		   Ta&4g&0.26140	&0.51821&0.11038\\
        Ni&2f&0.75&0.53097&	0.25\\
        Se(1)&4g&0.74288&0.43302&0.04885\\
        Se(2)&4g&0.74337&0.60868&0.13921\\
        Se(3)&2f&0.25&0.45075&0.25\\
			\hline
		\end{tabular}
	\label{table1}
\end{table}

We then evaluate the feasibility of exfoliating a monolayer from the bulk using the exfoliation energy, which is defined as $\Delta E_{\rm f}$=($E_{\rm {mono}}$ $-$ $\frac{1}{2}$$E_{\rm {bulk}}$)/$S$. Here $E_{\rm {mono}}$ and $E_{\rm {bulk}}$ correspond to the total energies per unit cell of the monolayer and the bulk [see Fig. 1(a)] Ta$_2$NiSe$_5$, respectively, and $S$ is the in-plane area. The calculated $\Delta E_{\rm f}$ is 26 meV/\AA$^2$. This value is just a little larger than 20 meV/\AA$^2$ of graphene\cite{Jung 2018} and well below the limit of 130$-$200 meV/\AA$^2$ for "potentially peelable" systems\cite{Mohanty 2020, Mounet 2018,Choudhary 2017}. Therefore, it is highly possible to prepare Ta$_2$NiSe$_5$ monolayer in an exfoliated manner.

%\subsection{\label{sec:level2}Electronic structure}

Figures 2(a) and 2(b) compare the band structures of bulk and monolayer Ta$_2$NiSe$_5$ by PBE. It can be seen that there is not much difference between the two, especially in terms of the gap size. This again indicates weak inter-layer interactions and also suggests an insignificant quantum size effect in 2D Ta$_2$NiSe$_5$. Indeed, although no monolayer was obtained, it was experimentally found that the gap in ultrathin Ta$_2$NiSe$_5$ is independent of the layer thickness (down to 5 layers)\cite{Kim 2016}. For the bulk, the lowest conduction band and the highest valence band almost intersect at the $\Gamma$ point, forming a $\sim$0 gap, which is consistent with previous study\cite{Ma 2022}. For the monolayer, these two bands exhibit "W" and "M" shaped dispersion along $A$-$\Gamma$-$Y_{2}$, opening a minimum gap of 36 meV. Typically, such a "W"-"M" feature signifies that a band inversion has occurred. When it is coupled with a SOC gap, it often means a non-trivial topology. However, two points are worth noting. On the one hand, the orbital-projected-band analysis in Fig. 2(c) shows that the states near the Fermi energy are dominated by Ta-5$d$ and Ni-3$d$. The former dominates the bottom conduction band, while both together dominate the top valence band. In general, all $d$-orbitals have the same parity, and band inversion between them does not lead to a non-trivial topology\cite{Lucking 2018, DongTEI 2023}. On the other hand, it is well known that PBE tends to grossly underestimate the gap of transition-metal compounds\cite{Tan 2019, LiuJ 2021, Qu 2023}. Given that it contains Ta and Ni, we first perform GW calculations to fix the gap underestimation before discussing the topological properties of the monolayer.

Figure 2(d) shows the quasi-particle bands. For comparison, we show the results without and with SOC separately. Three points are worth noting here. First, the gap increases significantly, to 210 and 266 meV in the case without and with SOC, respectively. Now the extremes of the valence and conduction bands are located at the same $k$ point, indicative of a narrow direct-gap semiconductor.

%fig02
\begin{figure}[htbp]
\includegraphics[width=0.95\columnwidth]{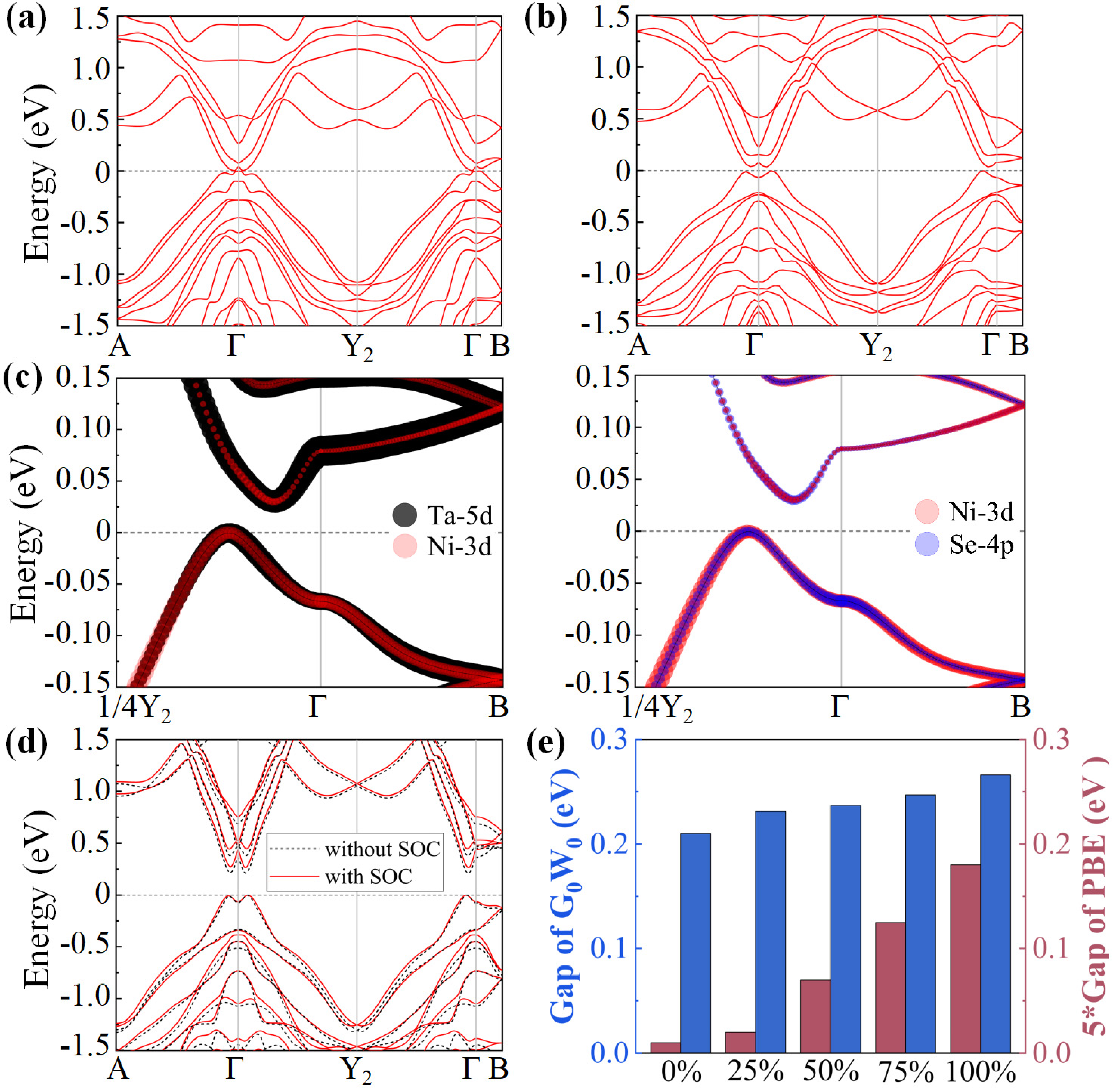}
\caption{\label{fig:fig2} Band structures of (a) bulk and (b) monolayer Ta$_2$NiSe$_5$ by PBE with SOC included. (c) Orbital-projected-band structures around the Fermi level corresponding to (b), with the line width proportional to the contribution weight. (d) Quasi-particle band structures of monolayer Ta$_2$NiSe$_5$ without and with SOC. In (a)-(d), the valence band maximum is set to zero energy. (e) Gap size as a function of SOC strength by GW (left blue axis) and PBE (right red axis), respectively. For clarity, the PBE gap is multiplied by 5. Note that there is a gap of $\sim$2 meV in the absence of SOC by PBE.}
\end{figure}

Second, unlike in the PBE where only the frontier states show "W"-"M" dispersion along the $A$-$\Gamma$-$Y_{2}$, the second-lowest conduction band also shows a clear "W"-shaped dispersion. This implies that band inversion may involve more than one conduction band and affect topological physics as well\cite{Lucking 2018}. In addition, a previous study suggests that non-trivial topology may exist in bulk Ta$_2$NiSe$_5$\cite{Ma 2022}. Given that both PBE and GW results show energy band inversion and SOC gap, we monitor the gap variation of monolayer Ta$_2$NiSe$_5$ by manually adjusting the SOC strength to check whether it has a non-trivial topology. Figure 2(e) gives the PBE and GW gaps as a function of SOC strength. It is clear that the gap increases monotonically with the SOC enhancement. There is no gap closure-reopening during this process, which indicates that the monolayer Ta$_2$NiSe$_5$ is topologically trivial.

%fig03
\begin{figure}[htbp]
\includegraphics[width=1\columnwidth]{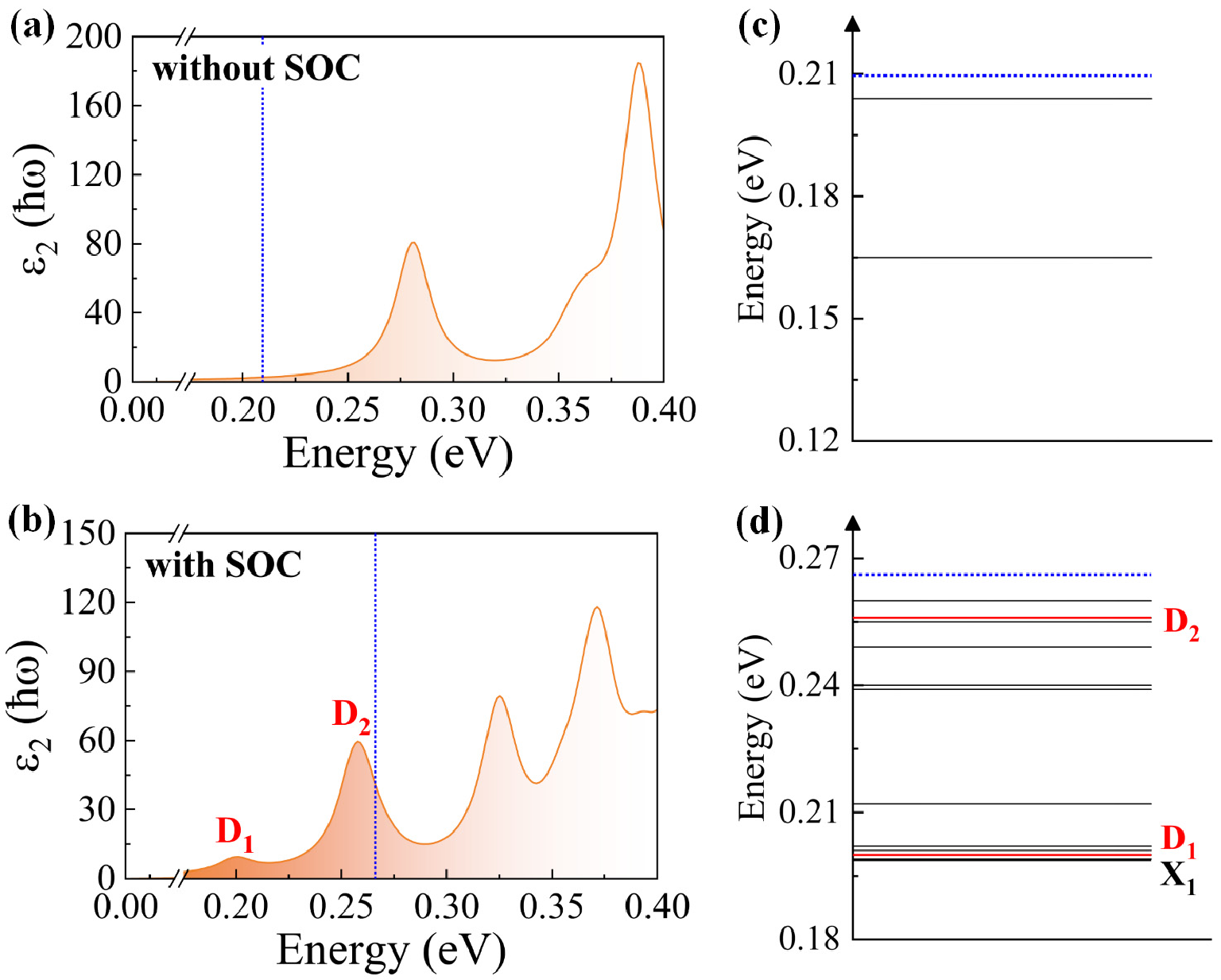}
\caption{\label{fig:fig3} (a)[b] Imaginary parts ($\varepsilon$$_2$) of the BSE dielectric function and (c)[d] low-energy exciton spectrum without[with] considering SOC. Blue dot lines denote the positions of quasi-particle gap. Each solid line in (c) and (d) represents an exciton state, with the lowest-energy exciton labelled $X_1$ and those corresponding to the absorption peaks $D_1$ and $D_2$ in (b) colored red.}
\end{figure}

Third, the Ta$_2$NiSe$_5$ crystal structure shown in Fig. 1(a) dictates an inherently in-plane anisotropy, which leads to significant anisotropy in mechanical properties\cite{Li 2024}, electrical transport, and optical response\cite{Qiao 2021}. It shows a fracture toughness anisotropy ratio of $\sim$3\cite{Li 2024}, an on/off current anisotropy ratio of $\sim$10, a mobility anisotropy ratio of 7.5, and an anisotropy ratio of 3.24 at 1064 nm laser illumination when used as polarization-sensitive photodetectors\cite{Qiao 2021}. As the monolayer maintains this structural anisotropy, it naturally exhibits some similar behaviors. Using the SOC bands in Fig. 2(d), we estimate the effective masses of the electrons (holes) along the chain direction ($\Gamma$-$Y_2$) and perpendicular to the chain direction ($\Gamma$-$B$) to be 0.03 (0.14) m$_0$ and 0.76 (0.36) m$_0$, respectively. Here m$_0$ is the mass of the free electron. Therefore, the anisotropy is very pronounced when the Ta$_2$NiSe$_5$ monolayer is electron-conducting and less pronounced when it is hole-conducting. Note that the band extremes lie on $A$-$\Gamma$-$Y_2$. When used as a semiconductor, if the Fermi energy is in the range ($-$66, 440) meV [see Fig. 2(d)], there is only one quasi-one-dimensional conducting channel along the direction of the trimetal chain. There is no anisotropy in this case. It is only when the Fermi energy enters the valence/conduction band through doping or gating beyond this range that the both channels along and perpendicular to the trimetal chain come into play and show anisotropy.

%\subsection{\label{sec:level2}Optical and excitonic properties}
Next, we turn to the optical and excitonic properties, as shown in Fig. 3. Imaginary parts of the dielectric function without and with SOC are given in Figs. 3(a) and 3(b), respectively. It can be seen that in the absence of SOC, there is no absorption peak inside the quasi-particle gap, hence corresponding to interband absorption. Its optical gap is now the same as the fundamental gap, which is 210 meV. When the SOC is considered, two absorption peaks appear below the quasi-particle gap at $D_1$ = 200 and $D_2$ = 256 meV, respectively. At this point, the optical gap is defined by exciton absorption, in contrast to the absence of SOC, where light absorption does not produce excitons. Although the optical gap of monolayer Ta$_2$NiSe$_5$ without and with SOC are comparable, both being $\sim$200 meV, it is clear that the two correspond to very different physics. Therefore, the SOC cannot be neglected when studying the optical properties of Ta$_2$NiSe$_5$, at least in the monolayer case. It is worth mentioning that, despite no experimental optical gap for monolayer Ta$_2$NiSe$_5$ so far, the 250 meV\cite{GuoT 2023} of the exfoliated multilayer is quite similar to our calculations.

To probe whether there are spontaneous excitons, we further provide the low-energy exciton spectrum, including both bright and dark, in Figs. 3(c) and 3(d), respectively. Without considering SOC, two dark excitons appear within the quasi-particle gap at 165 and 204 meV, respectively. Considering SOC, a number of excitons appear. The two bright excitons corresponding to the $D_1$ and $D_2$ peaks of Fig. 3(b) are marked with red lines in Fig. 3(d). We note that the lowest-energy exciton (dubbed as $X_1$) is a dark state with an energy only 1 meV lower than that of $D_1$. Because all exciton energies are positive, there are no spontaneous excitons in the Ta$_2$NiSe$_5$ monolayer.

The $X_1$ and $D_1$ excitons are almost degenerate in energy but have very different optical activities. To deepen the understanding, we write the exciton wave-function as a linear combination of the electron-hole pair states
\begin{equation}\label{(1)}
\Psi_{\textbf{q}}(r_h, r_e) = \sum_{vc\textbf{k}}A^\textbf{q}_{vc\textbf{k}}\psi_{v,\textbf{k}}(r_h)\psi^{*}_{c,\textbf{k}+\textbf{q}}(r_e),
\end{equation}
where $\psi_{v,\textbf{k}}(r_h)$ and $\psi_{c,\textbf{k}}(r_e)$ are the wave-functions of hole and electron, respectively.
In this way, we can define the physical quantity, $\zeta^\textbf{q}_{v(c)\textbf{k}}=\sum_{c(v)}|A^\textbf{q}_{vc\textbf{k}}|$$^2$ [with the summarization over the conduction (valence) band index, and \textbf{q} = 0 here], to visually inspect the relative contribution of each electron-hole pair in \textbf{k}-space to the very exciton eigenstate.

%fig04
\begin{figure}[htbp]
\includegraphics[width=1\columnwidth]{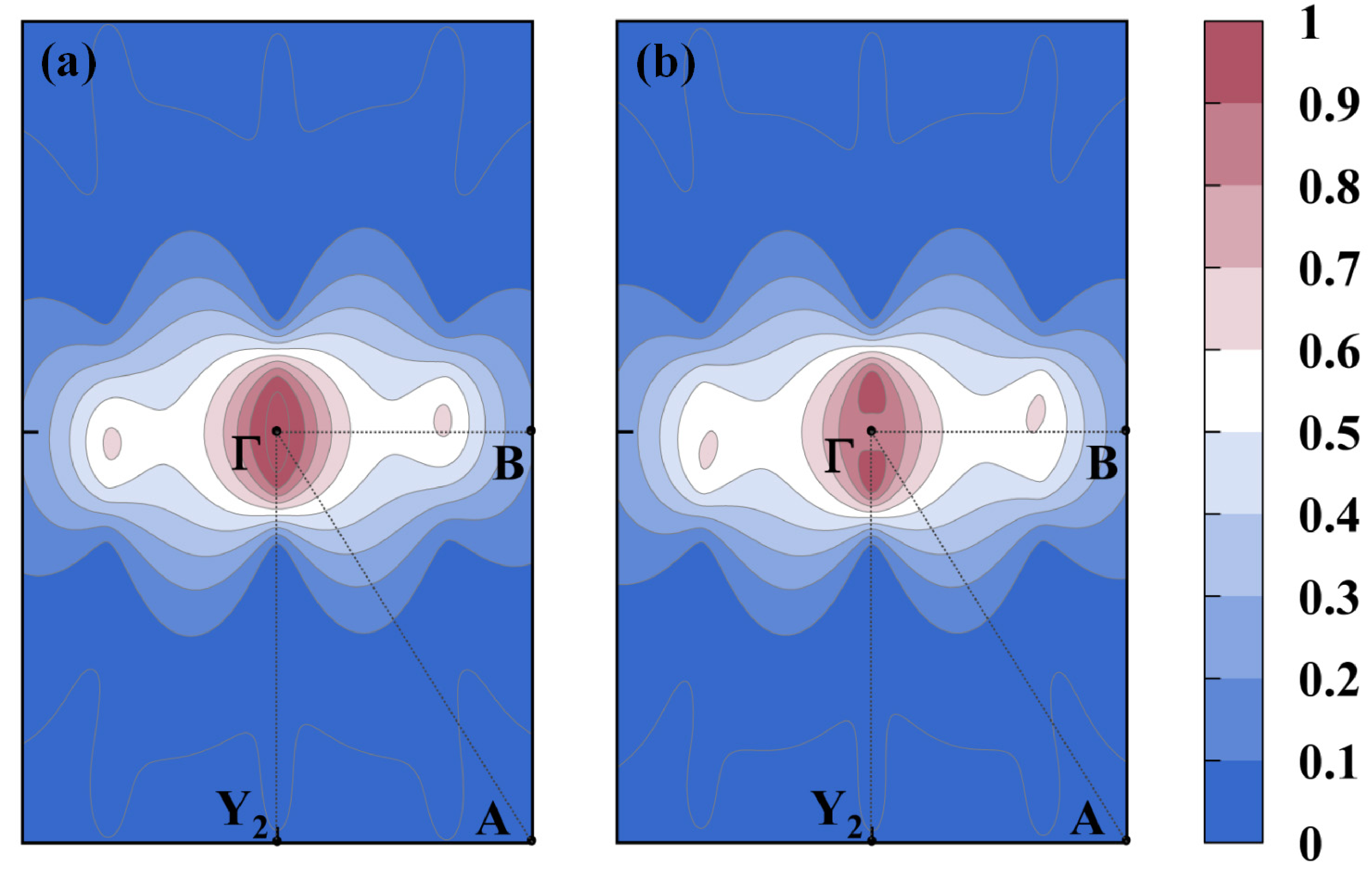}
\caption{\label{fig:fig4} Maps of $k$-point specific contribution to the (a) $X_1$ dark and (b) $D_1$ bright exciton wave-function modulus in the Brillouin zone. The density ($\zeta^\textbf{q}_{v(c)\textbf{k}}$) has been normalized by choosing the maximum value to be unity.}
\end{figure}

Figures 4(a) and 4(b) show the BSE fat-band structures in the reciprocal-space for $X_1$ and $D_1$ excitons, respectively. Obviously, the $X_1$-exciton spreads almost uniformly around the $\Gamma$-point, displaying a nonlocal feature. On the contrary the $D_1$-exciton is concentrated at two $k$ points on the $\Gamma$-$Y_2$, which happen to be the extremes of the "M"- and "W"-shaped bands. In this sense, the $D_1$-exciton is of the conventional Wannier-Mott type. Its binding energy is well defined by the difference between the corresponding single-particle gap of 266 meV and the excitation energy of 200 meV. The binding energy of 66 meV is about 1/4 of the quasi-particle gap, which is consistent with the unique scaling of 2D semiconductors\cite{Jiang 2017}. On the other hand, the nonlocal $X_1$-exciton does not have a well-defined single-particle gap, and hence no well-defined binding energy.

%\section{\label{sec:level1}Conclusions}
To summarize, our first-principles calculations coupled with Bethe-Salpeter equation show the experimental feasibility of exfoliating Ta$_2$NiSe$_5$ monolayer directly from its layered bulk counterpart. It maintains the geometry in the bulk to a large extent and is dynamically stable. The monolayer has fundamental and optical gaps of 266 and 200 meV, respectively, as well as an exciton binding energy of 66 meV. While spin-orbit coupling has a significant effect on excitons, for the gap it only plays a role in enlarging the gap by about one-third. In view of these, we conclude that monolayer Ta$_2$NiSe$_5$ is a common narrow-gap semiconductor with neither spontaneous excitons nor non-trivial topology.

%\begin{acknowledgments}
This work was supported by the Ministry of Science and Technology of China (Grant Nos. 2023YFA1406400 and 2020YFA0308800) and the National Natural Science Foundation of China (Grant No. 12074034).
%\end{acknowledgments}

\section*{Author declarations}
\subsection*{Conflict of Interest}
The authors have no conflicts to disclose.

\subsection*{Author contributions}
\textbf{Miaomiao Guo}: Investigation (lead); Software (lead); Writing-original draft (lead).
\textbf{Yuanchang Li}: Conceptualization (lead); Funding acquisition (lead); Supervision (lead); Writing-review \& editing (lead).

\section*{Data Availability Statement}

The data that support the findings of this study are available from the corresponding authors upon reasonable request.

\end{document}